\documentclass{icrc}

\usepackage{times}
\usepackage{graphicx}

\usepackage{makeidx}   % allows index generation
\usepackage{multicol}  % used for the two-column index
\usepackage{cropmark} % cropmarks for pages without
\usepackage{physprbb}  % centered layout of diverse elements, etc.
\def\lesssim{\mathrel{\hbox{\rlap{\hbox{\lower4pt\hbox{$\sim$}}}\hbox{$<$}}}}
\def\gtrsim{\mathrel{\hbox{\rlap{\hbox{\lower4pt\hbox{$\sim$}}}\hbox{$>$}}}}
\newcommand{\ffrac}[2]{\left( \frac{#1}{#2} \right)}

\begin{document}
\title{Ultra High Energy Neutrino-Relic Neutrino Interactions In Dark Halos
to Solve Infrared-Tev And GZK Cut-Off}
\author{D. Fargion, M.Grossi, P.G. De Sanctis Lucentini}

\affil{Physics Department and INFN, Rome University 1,
     Pl.A.Moro 2, 00185, Rome, Italy}
%     2 Physics Dept.New York University, N.Y., USA}

%\correspondence{daniele.fargion@roma1.infn.it}

\firstpage{1} \pubyear{2001}

\maketitle

\begin{abstract}
Ultra High Energy Neutrino scattering on Relic Light Neutrinos in
Dark Galactic or Local Group lead to Z and WW,ZZ showering : the
nucleon component of the shower may overcome the GZK cut-off
while the electro-magnetic tail at TeVs up to EeVs energy may
solve the Infrared-TeV cut-off in a natural way. Different Gamma
TeV puzzles may find a solution within this scenario: new
predictions on UHECR spectra in future data are derived.
\end{abstract}

\section{Introduction: GZK and Infrared-TeV cut offs}
The Ultra High Energy (UHE) neutrinos above GZK cut-off energies
may interact with relic cosmic ones leading (via $Z \, , \, W^+
W^- \, , \, ZZ $ productions) to a rich and complex chain both in
hadronic and electro-magnetic showers. The hadronic secondaries
may be source of extra-galactic Ultra High Energy (UHE) cosmic
rays (UHECR) originated at distances well above GZK($\gtrsim
10\,Mpc$) cut-off. The electro-magnetic secondaries either from
gauge bosons' decays in flight but mainly by lepton pair
productions ($\nu \nu \rightarrow e \tau$, $ e \mu$, $ \mu
\tau$), may pollute by synchrotron radiation showering first EeV,
and  PeV as well as  TeV gamma spectra. The survival TeV photons
may break and overcome the severe infrared - TeV cut-offs as
possibly observed in Mrk 501, in Mrk 421 and in possible GRB90417
TeV clustered photons.
 As a frame-work let us remind that modern
astro-particle physics face the old standing problem of dark
matter nature in galactic up to cosmic scales. Neutrino with a
light mass may play a key role in solving at least partially the
puzzle within a hot-cold dark matter (HCDM) scenario. Moreover,
at the edge of highest energy astrophysics, the main open
question regards the nature of highest (Ultra High Energy, UHE)
cosmic rays above the Greisen Zatsepin Kuzmin cut-off ($\gtrsim 4
\cdot 10^{19}\,eV$). These rare events probably ejected by
blazars AGN, or QSRs in standard scenario, should not come from
large distances because of the electromagnetic "dragging
friction" of cosmic 2.75 K BBR and of the diffused radio
backgrounds. According to Greisen, Zatsepin and Kuzmin (1966)
proton and nucleon mean free path at E $> 5 \cdot 10^{19} \,EeV$
is less than 30 $Mpc$, while gamma rays at same energies have
even shorter interaction length ($10 \,Mpc$) due to microwave and
radio background (Prothoroe 1997). Nevertheless at nearby
distances ($\lesssim 10 \div 20 \, Mpc$), these powerful sources
(AGN, Quasars) suspected to be the unique source able to eject
such UHECRs, are rare and in general absent in the few (tens)
UHECR arrival directions.  Strong and coherent magnetic fields,
able to bend UHECR from nearer AGN sources the UHECR (proton,
nuclei) have been proposed either by galactic and extragalactic
origin. The needed magnetic coherent lengths and strength are not
compatible with known data. Moreover the absence of any un-bent
UHECR neutrons at GZK energies tracing the original source makes
the galactic or extra-galactic magnetic field solution less
acceptable. Topological defects (TD) clustering in the dark halo
were recently suggested to be the UHECR solution, but  growing
evidences for non homogeneous UHECR clustering in their arrival
directions (in doublets and triplets), are standing in favor of
compact UHECR sources (and not for any diffused dark TD clouds).
Moreover recent identification of UHECR with nearest Blazars at
redshift distances greater than allowable by GZK ones made more
than ever necessary the UHECR propagation from cosmic distances
(Tinyakov, Tkachev 2001). To escape this arguments there have
been even recent suggestions and speculations (Blasi 2000) for an
unexpected population of 500 compact dark clouds of $10^8
M_{\odot}$ each of TD clusters, nevertheless uncorrelated to
visible galactic halo and disk.   Therefore the solution of UHECR
puzzle based on primary Extreme High Energy (EHE) neutrino beams
(from far AGN) at $E_{\nu} > 10^{21}$ eV and their undisturbed
propagation up to the interaction on relic light $\nu$ in dark
galactic halo (Fargion,Salis 1997; Fargion, Mele, Salis 1999,
Weiler 1999, Yoshida et all 1998) is still a favorite option. If
relic neutrinos have a mass around few eVs to cluster in galactic
halos, the collisions with EHE neutrinos determine high energy
particle cascades which could contribute or dominate the observed
UHECR flux above $5 \cdot 10^{19} \,eV$. Indeed the possibility
that neutrino share a little mass has been reinforced by
Super-Kamiokande evidence for atmospheric neutrino anomaly via
$\nu_{\mu} \leftrightarrow \nu_{\tau}$ oscillation. Consequently
there are two extreme scenario for hot dark halos: either
$\nu_{\mu}\, , \, \nu_{\tau}$ are both extremely light
($m_{\nu_{\mu}} \sim m_{\nu_{\tau}} \sim \sqrt{\Delta m} \sim
0.07 \, eV$) and therefore hot dark neutrino halo is very wide
and spread out to local group clustering sizes (increasing the
radius but loosing in the neutrino density clustering contrast),
or $\nu_{\mu}, \nu_{\tau}$ have degenerated (eVs - 10 eV masses)
split by a very tiny different value. \\ In the first case the Z
peak $\nu$ - $\nu_r$ interaction (Fargion Salis 1997,Fargion,Mele
Salis 1999, T.Weiler 1999,Yoshida et all 1998) will be the
favorite one while in the second case a $\nu \nu$ interaction via
$Z \rightarrow W^+ W^-$ (Fargion,Mele,Salis 1999) and $\nu \nu
\rightarrow ZZ$ (D.Fargion, M.Grossi, P.G.De Sanctis Lucentini,
C.Di Troia, R.V.Konoplich 2001) will be the dominant one.

%%%%%%%%%%%%%%%   Figure 1 Ex 01  %%%%%%%%%%%%%%%%%   FFFFFFFFFFFFFFFFFFFFFFFFFFFFFFFFFFFFFF
\begin{figure}[h]
\begin{center}
 \includegraphics[width=0.45\textwidth] {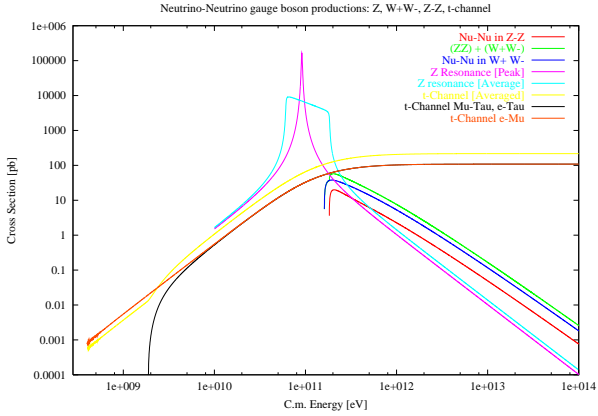}
\end{center}
  \caption{The  $\nu \bar{\nu} \rightarrow  Z,W^+ W^-,ZZ
  ,T$-channel,  cross sections as a function of the center of mass energy in $\nu \nu$.
   These cross-sections are estimated also in average (Z) as well for each possible
   t-channel lepton pairs. The averaged t-channel averaged the multiplicity of flavours
    pairs ${\nu}_{i}$, $ \bar{\nu}_{j}$ respect to neutrino
    pair annihilations into Z neutral boson.}
 \label{fig:boxed_graphic 1}
\end{figure}
%%%%%%%%%%%%%%%   Figure 1 END   %%%%%%%%%%%%%%%%%   FFFFFFFFFFFFFFFFFFFFFFFFFFFFFFFFFFFFFF

The hadronic tail of the Z or $W^+ W^-$ cascade might be the
source of final $p, \bar{p}, n, \bar{n}$ able to explain UHECR
events observed by Fly's Eye and AGASA . However the same $\nu
\nu_r$ interactions are source of Z and W decaying into pions as
well as UHE lepton couples that are source of UHE gamma and
electron pairs. Their  average energy deposition for both gauge
bosons decay is split as in short summary and detailed table
below.

\begin{table}[h]
\begin{center}
\begin{tabular}{|c|c|c|c|}
\hline
   & Z & $W^+ W^-$ & t-channel\\ \hline
  $\nu$ & 58 \% & 55 \% & 47 \% \\ \hline
    $\gamma$
& 21 \% & 21 \% & 4 \% \\ \hline
    $e^+ e^-$ & 16 \% & 19 \% & 49 \% \\ \hline
  $p$ & 5 \% & 5 \% & - \\ \hline
\end{tabular}
\end{center}
\end{table}

\begin{figure}[h]
\begin{center}
 \includegraphics[width=0.45\textwidth] {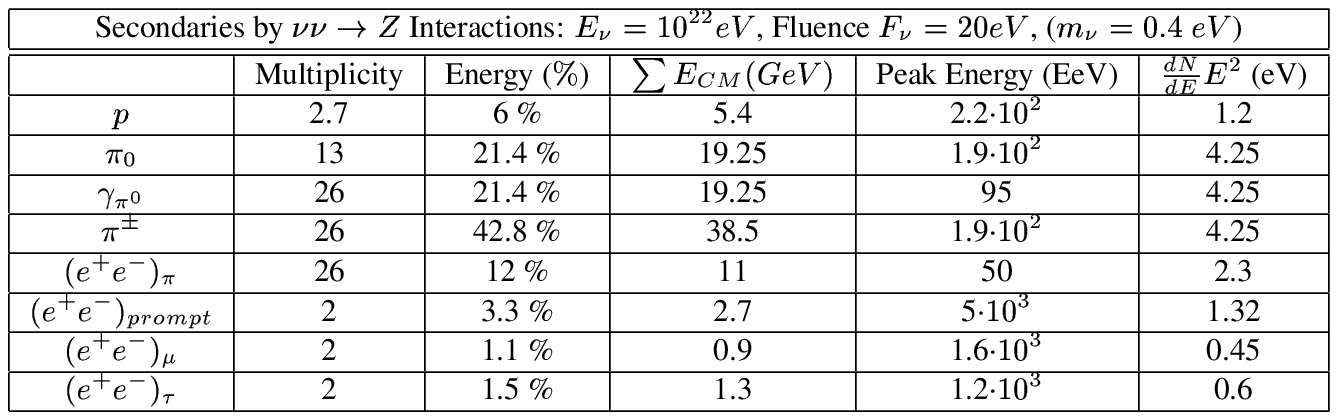}
\end{center}
  \caption{Total Energy percentage distribution  into
  protons, neutral and charged pions and consequent gamma, electron
pair particles both from hadronic and prompt leptonic Z, $WW,ZZ$
productions and t-channels. We also calculated the elecro-magnetic
contribution due to the t-channel $\nu_i \nu_j$ interactions. We
used LEP data for Z decay and considered W decay roughly in the
same way as Z one. We assumed that an average number of 37
particles is produced during a Z (W) hadronic decay. The number
of prompt pions both charged (18) and neutral (9), in the
hadronic decay is increased by 8 and 4 respectively due to the
decay of $K^0$, $K^{\pm}$, $\rho$, $\omega$, and $\eta$
particles. We assumed that the most energetic neutrinos produced
in the hadronic decay mainly come from charged pion decay.  Their
number is roughly three times the number of $\pi$'s. UHE photons
are mainly relics of neutral pions. Most of the $\gamma$
radiation will be degraded around PeV energies by $\gamma \gamma$
pair production with cosmic 2.75 K BBR, or with cosmic radio
background. The electron pairs instead, are mainly relics of
charged pions and will rapidly lose energies into synchrotron
radiation. Also prompt electron pairs (or leptonic secondaries)
are considered. The consequent energy injection at each cosmic
ray component and their showering spectra derived by boosted Z,WW
decay in flight is shon in Figures below.} \label{1}
\end{figure}
Gamma photons at energies $E_{\gamma} \simeq 10^{20}$ - $10^{19}
\,eV$ freely propagate through galactic halo scales (hundreds of
kpc) and could contribute to the extreme edge of cosmic ray
spectrum (Yoshida et all 1998). The ratio of the final fluxes of
nucleons, $\Phi_p$ over the corresponding electromagnetism flux
$\Phi_{em}$ ratio is nearly 1 / 20. Moreover if one considers the
six pairs $\nu_e \bar{\nu_e}$,$\nu_{\mu} \bar{\nu_{\mu}}$,
$\nu_{\tau} \bar{\nu_{\tau}}$ (and their hermite conjugate)
t-channel interactions leading to discrete nearly monochromatic
leptons at high energy, with no hadronic relics (as $p,
\bar{p}$). This additional injection favors the electro-magnetic
flux $\Phi_{em}$ over the corresponding nucleon one $\Phi_p$ by a
factor 2 leading to $\frac{\Phi_p}{\Phi_{em}} \sim \frac{1}{40}$.
We shall focus on Z, and WW, channels and their consequent
electro-magnetic showering . These cascade will pile up into TeVs
and later in GeVs while the previous one will double the
electromagnetic shower mostly directly into the MeV band.
\begin{figure}[h]
\begin{center}
 \includegraphics[width=0.45\textwidth] {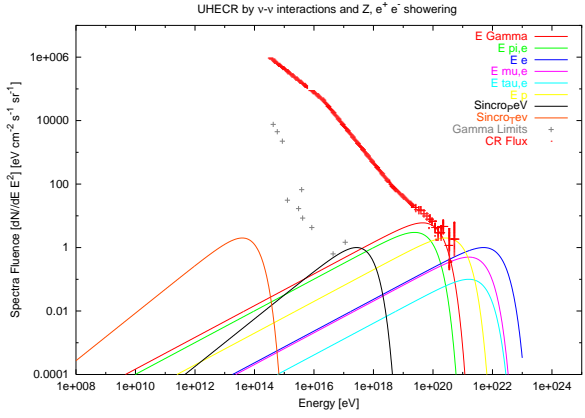}
\end{center}
\caption{Z decay Showering into nucleons,gamma and electron pairs
over observed cosmic ray data and TeV gamma bounds. The sincro (T
TeV, P PeV)  gamma are degraded synchrotron radiation by UHE
electron pairs secondaries  in galactic magnetic fields. Here we
assumed an incoming UHE neutrino at $10^{22} $ eV and a relic
neutrino mass at 0.4 eV. The UHE Gamma are relics of neutral
pions decay, $E_p$ labels the nucleons,$E_e$ the UHE prompt
electron pairs,$E_{\mu}$,$E_{\tau}$ by Z decay.}
\label{fig:boxed_graphic 7}
\end{figure}
%%%%%%%%%%%%%%%   Figure 7END   %%%%%%%%%%%%%%%%%   FFFFFFFFFFFFFFFFFFFFFFFFFFFFFFFFFFFFFF

%%%%%%%%%%%%%%%   Figure 8  WWWWWWWW-t-channel  %%%%%%%%%%%%%%%%%   FFFFFFFFFFFFFFFFFFFFFFFFFFFFFFFFFFFFFF
\begin{figure}[h]
\begin{center}
 \includegraphics[width=0.45\textwidth] {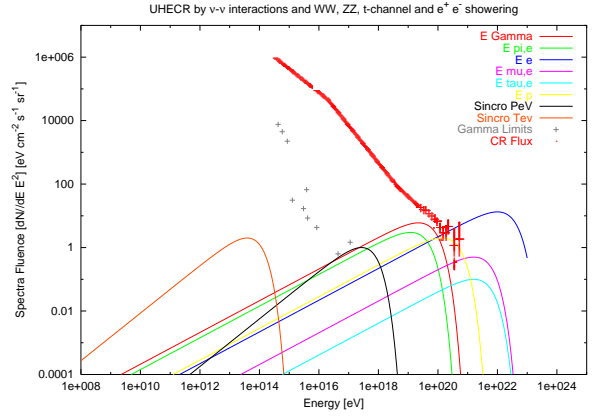}
\end{center}
\caption{WW and ZZ decay Showering as well as the t-channel UHE
leptons decaying into nucleons,gamma and electron pairs over
written on observed cosmic ray data and TeV gamma bounds. The
sincro (T TeV, P PeV) gamma are degraded synchrotron radiation by
UHE electron pairs secondaries in galactic magnetic fields. Note
the electron pair excess at highest energies. Their presence due
to t-channel showering should be masked by radio scattering and
magnetic screening . Here we assumed an incoming UHE neutrino at
$10^{22} $ eV and a relic neutrino mass at 0.4 eV. The UHE Gamma
are relics of neutral pions decay, $E_p$ labels the
nucleons,$E_e$ the UHE prompt electron pairs,$E_{\mu}$,$E_{\tau}$
by Z decay.}
 \label{fig:boxed_graphic 8}
\end{figure}
It is important to summarize the shower chain looking first to the
EHE component and than following their consequent migration in
lower energies.

\section{The electromagnetic imprint of UHE neutrino scattering in the halo}

Extra-galactic neutrino cosmic rays are free to move on cosmic
distances up our galactic halo without constraint on their mean
free path, because the interaction length with cosmic background
neutrinos is greater than the actual Hubble distance . A Hot Dark
Matter galactic halo model with relic light neutrinos (primarily
the heaviest $\nu_{\tau}$ or $ \nu_{\mu} $), acts as a target for
the high energy neutrino beams. The relic number density and the
halo size are large enough to allow the $\nu \nu_{relic}$
interaction . As a consequence high energy particle showers are
produced in the galactic halo, overcoming the GZK cut-off. There
are three main processes that can occur in the interaction  of
EHE and relic neutrinos. The $\nu \nu_r\rightarrow Z \rightarrow
\, $
 annihilation at the Z resonance; the second ($\nu_{\mu} \bar{\nu_{\mu}}
\rightarrow W^+ W^- \rightarrow$  hadrons, electrons, photons)
through  W pair production.
 The third $\nu_e$ - $\bar{\nu_{\mu}}$, $\nu_e$ -
$\bar{\nu_{\tau}}$, $\nu_{\mu}$ - $\bar{\nu_{\tau}}$ and its
hermite conjugate interactions of different flavor neutrinos
mediated in the  $t$ - channel by the W exchange (i.e. $\nu_{\mu}
\bar{\nu_{\tau_r}} \rightarrow \mu^- \tau^+ $). These reactions
are sources of prompt and secondary UHE electrons as well as
photons resulting by hadronic $\tau$ decay.
%\section{The prediction of the UHE particles spectra from W and Z decay}
%Let us examine the destiny of UHE primary particles (nucleons,
%electrons and photons) ($E_e \lesssim 10^{21}\,eV$) produced
%after hadronic or leptonic W decay.
As we already noticed in the introduction, we'll assume that the
nucleons, electrons and photons spectra (coming from W or Z
decay) after $\nu \nu$ scattering in the halo, follow a power law
that in the center of mass system is $\frac{dN^*}{dE^* dt^*}
\simeq E^{* - \alpha}$ where $\alpha \sim 1.5$. This assumption
is based on detailed Monte Carlo simulation of a heavy fourth
generation neutrino annihilations (Golubkov,Konoplich 1998)
(Fargion et all 1999.2001) and with the model of quark - hadron
fragmentation spectrum suggested by Hill (1983). In order to
determine the shape of the particle spectrum in the laboratory
frame, we have to introduce the Lorentz relativistic
transformations from the center of mass system to the laboratory
system. The number of particles is clearly a relativistic
invariant $dN_{lab} = dN^*$, while the relation between the two
time intervals is $dt_{lab} = \gamma dt^*$, the energy changes
like $ \epsilon_{lab} = \gamma \epsilon^* (1 + \beta \cos
\theta^*) = \epsilon^* \gamma^{-1}(1 - \beta \cos \theta)^{-1}$,
and finally the solid angle in the laboratory frame of reference
becomes $d\Omega_{lab} =\gamma^{2} d\Omega^*  (1 - \beta \cos
\theta )^2$. Substituting these relations one obtains

\begin{displaymath}
\ffrac{dN}{d\epsilon dt
d\Omega}_{lab}=\frac{dN_{*}}{d\epsilon_{*} dt_{*}
d\Omega_{*}}\gamma^{-2}(1 - \beta \cos \theta)^{-1}=
\end{displaymath}

\begin{displaymath}
=\frac{\epsilon^{-\alpha}_{*} \gamma^{-2}}{4\pi} \cdot (1 - \beta
\cos \theta)^{-1}
\end{displaymath}

\begin{equation}
{{\left( \frac{dN}{d\epsilon dt d\Omega} \right)_{lab} =
 \frac{\epsilon^{-\alpha} \; \gamma^{-\alpha-2}} {4 \pi}} (1 - \beta \cos \theta)^{-\alpha-1}
 }
\end{equation}

and integrating on $\theta$ (omitting the lab notation) one loses
the spectrum dependence on the angle.

%\begin{equation}
%\left( \frac{dN}{d\epsilon dt d\Omega} \right)_{lab} \propto
%\epsilon^{-\alpha} \gamma_{Z (W)}^{ \alpha} \sim \epsilon^{-
%\frac{\alpha}{2}} \sim \epsilon^{- 0.75}.
%\end{equation}

The consequent fluence derived by the solid angle integral is:
% Long version
%\begin{equation}
% \frac{dN}{d\epsilon dt} \epsilon^{2}=
% \frac{\epsilon^{-\alpha+2} \; \gamma^{\alpha-2}} {2 \beta \alpha}}
% [(1 + \beta)^{\alpha} -
%  \frac{1} {[(1 + \beta)\gamma^2]^{\alpha}}}] \simeq
% \frac{\epsilon^{-\alpha+2} \; \gamma^{\alpha-2}} {\alpha}}
%\end{equation}

\begin{equation}
{ \frac{dN}{d\epsilon dt} \epsilon^{2} \simeq
 \frac{2^{\alpha-1}\epsilon^{-\alpha+2} \; \gamma^{\alpha-2}} {\alpha}}
\end{equation}

There are two extreme case to be considered: the case where the
interaction occur at Z peak resonance and therefore the center of
mass Lorents factor $\gamma$ is "frozen" at a given value (eq.1)
and the case (WW,ZZ pair channel) where all energies are
allowable  and $\gamma$ is proportional to $\epsilon^{1/2}$.
%; the latter case will be discussed in detail elsewhere.
 Here we focus only on Z peak resonance. The consequent fluence spectra
 $\frac{dN}{d\epsilon dt}\epsilon^{2}$, as above, is proportional to $\epsilon^{-\alpha +2}$. Because $\alpha$ is
nearly $1.5$ all the consequent secondary particles will also show
a spectra proportional to $\epsilon^{1/2}$ following a normalized
energies shown in Tab.2, as shown in figures above. In the latter
case (WW,ZZ pair channel), the relativistic boost reflects on the
spectrum of the secondary particles, and the spectra power law
becomes $\propto \epsilon^{\alpha/2 +1}=\epsilon^{0.25}$. These
channels will be studied in details elsewhere. In Fig. 1 we show
the spectrum of protons, photons and electrons coming from Z
hadronic and leptonic decay assuming a nominal primary CR energy
flux $\sim 20 eV s^{-1} sr^{-1} cm^{-2}$, due to the total $\nu
\bar{\nu}$ scattering at GZK energies. Let us remind that we
assume an interaction probability of $\sim 1 \%$ and a
corresponding UHE incoming neutrino energy $\sim 2000 eV s^{-1}
sr^{-1} cm^{-2}$ near but below present $UHE$ neutrino flux bound.

%\subsection{Electron synchrotron radiation in the galactic halo}

%As we underlined in the last section, electrons (and positrons)
%are produced either in hadronic or in leptonic Z and W decay. The
%two different channels lead to different electron energies  and
%fluxes. The less probable $Z \rightarrow e^+ e^- (\mu^+ \mu^- \, ,
%\, \tau^+ \tau^-)$ or $W \rightarrow e \nu_e (\mu \nu_{\mu} \, ,
%\, ...)$ channels give lightest charged leptons (see Table 3, 4)
%with energies closer to the W boson ones ($E_e = E_{Z(W)}/2$ for
%prompt electrons and $E_e^{\mu} (E_e^{\tau}) = E_{Z(W)}/6$ for
%electrons by muon and tau decay). While the electrons (positrons)
%by the hadronic channel (see Table 1), even if more abundant (only
%seventeen electrons and positrons on average by charged pion
%decay), have just few per cent of the initial gauge boson
%energy.\\

 As it is shown in Table 2 and Figures above, the electron
(positron) energies by $\pi^{\pm}$ decays is around $E_e \sim 2
\cdot 10^{19} \, eV$ for an initial $E_Z \sim 10^{22} \, eV $ (
and $E_{\nu} \sim 10^{22} \, eV $). Such electron pairs
interacting with the galactic magnetic field ($B_G \simeq 10^{-6}
\,G $) will lead, to direct TeV photons:
\begin{displaymath}
  E_{\gamma}^{sync} \sim \gamma^2 \left( \frac{eB}{2\pi m_e } \right)
   \sim
\end{displaymath}
\begin{equation}\label{4}
  \sim 27.2 \left( \frac{E_e}{2 \cdot10^{19}
  \,eV} \right)^2 \left( \frac{m_{\nu}}{0.4 \, eV} \right)^{-2} \left( \frac{B}{\mu G} \right)\,TeV.
\end{equation}
The spectrum of these photons is characterized by a power of law
$dN /dE dT \sim E^{-(\alpha + 1)/2} \sim E^{-1.25}$ where
$\alpha$ is the power law of the electron spectrum, and it is
showed in Figures above. As regards the prompt electrons at
higher energy ($E_e \simeq 10^{21}\, eV$), their interactions
with the galactic magnetic field is source of another kind of
emission around tens of PeV energies, as is given by :
\begin{equation}\label{2}
  E^{sync}_{\gamma}  \sim
  6.8 \cdot 10^{16} \left( \frac{E_e}{10^{21}\,eV} \right)^2
  \left( \frac{m_{\nu}}{0.4 \, eV} \right)^{-2} \left( \frac{B}{\mu G}
  \right) \, eV
\end{equation}
The corresponding energy loss length instead is (Kalashev,
Kuzmin, Semikoz 2000)
\begin{equation}\label{3}
\left( \frac{1}{E} \frac{dE}{dt} \right)^{-1} = 3.8 \times \left(
\frac{E}{10^{15}} \right)^{-1} \left( \frac{B}{10^{-6} G}
\right)^{-2} \, kpc.
\end{equation}
that for $10^{21}$ electron reduces to $\sim 10^{-3} pc$. Again
one has the same power law characteristic of a synchrotron
spectrum with index $E^{-(\alpha + 1 / 2)} \sim E^{-1.25}$.
 Gammas at $10^{16} \div 10^{17}$ eV scatters onto
low-energy photons from isotropic cosmic background ($\gamma + BBR
\rightarrow e^+ e^-$) converting their energy in electron pair.
 The expression of the pair production cross-section is
is
\begin{equation}
\sigma (s) = \frac{1}{2} \pi r_0^2 (1 - v^2) [ (3 - v^4) \ln
\frac{1 + v}{1 - v} - 2 v (2 - v^2) ]
\end{equation}
where $v = (1 - 4m_e^2 / s)^{1/2}$,  $s = 2 E_{\gamma} \epsilon (1
- \cos \theta)$ is the square energy in the center of mass frame,
$\epsilon$ is the target photon energy, $r_0 $  is the classic
electron radius, with a peak cross section value at
\[ \frac{4}{137}\times \frac{3}{8\pi} \sigma_T \ln 183 = 1.2 \times
10^{-26} \,cm^2 \] Because the corresponding attenuation length
due to the interactions with the microwave background is around
ten kpc , the extension of the halo plays a fundamental role in
order to make this mechanism efficient or not. As is shown in
Fig.3-4, the contribution to tens of PeV gamma signals by Z (or
W) hadronic decay, could be compatible with actual experimental
limits fixed by CASA-MIA detector on such a range of energies.
Considering a halo extension $l_{halo} \gtrsim 100 kpc$, the
secondary electron pair creation becomes efficient, leading to a
suppression of the tens of PeV signal. So electrons at $E_e \sim
3.5 \cdot 10^{16} \,eV$ loose again energy through additional
synchrotron radiation, with maximum around
\begin{equation}\label{3b}
  E_{\gamma}^{sync} \sim 79 \left( \frac{E_e}{10^{21}
  \,eV} \right)^4 \left( \frac{m_{\nu}}{0.4 \, eV} \right)^{-4}
   \left( \frac{B}{\mu G} \right)^3 \, MeV.
\end{equation}
Anyway this signal is not able to pollute sensibly the MeV-GeV
background, and its intensity is restricted beneath EGRET
detecting capacities.
%It has to underline that an almost similar result could occur for
%a W decay in a $\nu \nu$ scattering out of the Z peak, with an
%initial $E_{\nu} \sim 2 \cdot 10^{22} \, eV$. The introduction of
%such a reaction anyway implies to consider as well the
%consequences of the scattering mediated in the t-channel (see
%section 3.3), whose production of extremely high energetic
%electrons at $E_e \sim 10^{22}$ lead to a very high bump in
%hundreds of PeV renge of energies.
%%%%%%%%%%%%%%%%%%%%%%%%%%%%%%%%%%%%%%%%%%%%%%%%%%%%%%%%%%%%%%%%%%%%%%%%%%%%%%%5
Gamma rays with energies up to 20 TeV have been observed by
terrestrial detector only by nearby sources like Mrk 501 (z =
0.033) or very recently by MrK 421. This is puzzling because the
extragalactic TeV spectrum should be, in principle, significantly
suppressed by the $\gamma$-rays interactions with the
extragalactic Infrared background, leading to electron pair
production and TeVs cut-off. The recent calibration and
determination of the infrared background by DIRBE and FIRAS on
COBE have infered severe constrains on TeV propagation. Indeed,
as noticed by Kifune (2000), and Protheroe and Meyer (2000) we may
face a severe infrared background - TeV gamma ray crisis. This
crisis imply a distance cut-off, incidentally, comparable to the
GZK one. Let us remind also an additional problem related to the
possible discover of tens of TeV counterparts of BATSE
GRB970417,observed by Milagrito (1999), being most GRBs very
possibly at cosmic edges, at distances well above the IR-TeV
cut-off ones. One may invoke unbelievable extreme hard intrinsic
spectra or exotic explanation as gamma ray superposition of
photons or sacrilegious  Lorentz invariance violation
(G.Amelino-Camelia 1998).
\section{Conclusions}
We suggest that the same UHE $\nu \nu_r$ interaction chain in
dark halos, while being able to solve the UHECR GZK puzzle, may at
the same time, solve the present IR-TeV  paradox.

\end{document}